\documentclass[12pt]{article}
\usepackage{amssymb,amsmath}
\begin{document}


\sloppy
\title
{\hfill{\normalsize\sf FIAN/TD/01-05}    \\
            \vspace{1cm}
{\Large Vector boson in constant electromagnetic field }
}
\author
 {
       A.I.Nikishov
          \thanks
             {E-mail: nikishov@lpi.ru}
  \\
               {\small \phantom{uuu}}
  \\
           {\it {\small} I.E.Tamm Department of Theoretical Physics,}
  \\
               {\it {\small} P.N.Lebedev Physical Institute,}
  \\
         {\it {\small} 117924, Leninsky Prospect 53, Moscow, Russia.}
 }
\date{28 Feb 2001}
\maketitle
\begin{abstract}
 The propagator and complete sets of in- and out-solutions
of wave equation, together with Bogoliubov coefficients, relating these
solutions, are obtained for vector $W$-boson (with gyromagnetic ratio
 $g=2$) in a
 constant electromagnetic field.
When only electric field is present the Bogoliubov coefficients are independent
of boson polarization and are the same as for scalar boson. When both electric
 and magnetic fields are present and collinear, the Bogoliubov coefficients for
 states with boson spin perpendicular to the field
are again the same as in scalar case. For $W^-$ spin parallel (antiparallel)
to the magnetic field the Bogoliubov  coefficients and contributions to the
imaginary part of the Lagrange function in one loop approximation are obtained
 from corresponding
expressions for scalar case by substitution $m^2\to m^2+2eH$ $(m^2\to m^2-2eH)$.
For gyromagnetic ratio $g=2$ the vector boson interaction with constant
electromagnetic field is described by the functions, which can be expected
by comparing wave functions for scalar and Dirac particle in constant
electromagnetic field.
\end{abstract}

\newpage

\section{Introduction}

 Vector bosons occupy an intermediate place between low spin particles
(with spin 0 and 1/2) and higher spin particles. As such they can share some
of the problems encountered in considering higher spin particle interaction with
 strong electromagnetic field. The most conspicuous feature of vector boson
interaction in case of $g=2$ is the appearance of tachionic modes in
 overcritical magnetic field. The ways to deal with this problem in the
framework of non-abelian theories are analyzed in [1]. But are there any
others? According to [2] just in case $g=2$ there are problems in treating
pair production by electric field, using the method of diagonalization of
Hamiltonian. This is surprising in view of successful calculation of the
Lagrange function of constant field in one loop approximation [3]. We calculate
pair production by constant field, using Bogoliubov coefficients (which contain
all the information about this process) and obtained, as expected, the results
in agreement with [3] and [4].

\section{Vector boson in constant electric field }

We assume $\eta_{\mu\nu}={\rm diag(-1,1,1,1)}$ and denote $e=\vert e\vert$. The wave
function  of $W^-$ boson ($g=2$) in a source-free space
(where $\partial_\mu F^{\mu\nu}=0$) satisfies the
equation [1, 5]
$$
 (-D_{\sigma}D^{\sigma}+m^2)\psi_{\mu}-2ieF_{\mu\nu}\psi^{\nu}=0 \eqno (1)
$$
and constraint
$$
D_{\mu}\psi^{\mu}=0,\quad D_{\mu}=\partial_{\mu}+ieA_{\mu}.\eqno(2)
$$
Taking the vector potential in the form $A_3=-Et,A_1=A_2=A_0=0$,
 we find from (1) that the
 equations for $\psi^1$ and $\psi^2$ are the same as in scalar case
$$
(-D^2+m^2)\psi^{1,2}=0.                  \eqno(3)
$$
For $\psi^3$ and $\psi^0$ we get from (1)
$$
(-D^2 +m^2)\psi^3-2ieE\psi^0=0,\quad
(-D^2+m^2)\psi^0-2ieE\psi^3=0.      \eqno(4)
$$
Introducing $\psi^{\pm}=\psi^0\pm\psi^3$ we rewrite (4) as follows
$$
(-D^2+m^2\mp2ieE)\psi^{\pm}=0, \eqno(5)
$$
i.e. it is obtainable from (3) by substitution $m^2\to m^2\mp2ieE$. We
 see that the wave function for vector boson can be obtained from corresponding
wave function for scalar boson by simple rules.

Now we shall do this. The {\it positive-frequency} in-solution for
(negatively charged) scalar boson we denote as ${}_+\psi_{\vec p}$.
Here $\vec p=(p_1,p_2,p_3)$; this index will be dropped in what follows.
 Then [6]
$$
{}_+\psi\propto D_{\nu}(\tau)\exp{i\vec p\vec x}.     \eqno(6)
$$
Here $D_{\nu}(\tau)$ is parabolic cylinder function [7] and
$$
\nu=\frac{i\lambda}{2}-\frac12,\quad \tau=-\sqrt{2eE}e^{-i\frac{\pi}{4}}%
(t-\frac{p_3}{eE}),\quad \lambda=\frac{m^2+p_1^2+p_2^2}{eE}.     \eqno(7)
$$
For vector boson we get
$$
{}_+\psi=\begin{bmatrix}
\psi^0\\
\psi^1\\
\psi^2\\
\psi^3
\end{bmatrix}
 =\begin{bmatrix}
c_+D_{\nu+1}(\tau)+c_-D_{\nu-1}(\tau)\\
\hspace{0.15cm} c_1D_{\nu}(\tau)\\
\hspace{0.15cm} c_2D_{\nu}(\tau)\\
c_+D_{\nu+1}(\tau)-c_-D_{\nu-1}(\tau)\\
\end{bmatrix}e^{i\vec p\cdot\vec x}.             \eqno(8)
$$
Here $\psi^1={}_+\psi^1,\quad \psi^2={}_+\psi^2$ and
\setcounter{equation}{8}
\begin{gather}
\psi^0\equiv{}_+\psi^0=\frac12({}_+\psi^++{}_+\psi^-),\quad
 \psi^3\equiv{}_+\psi^3=
\frac12({}_+\psi^+-{}_+\psi^-),\\     \notag
 {}_+\psi^{\pm}=2c_{\pm}D_{\nu\pm1}%
e^{i\vec p\cdot\vec x}.                                        
\end{gather}
 $D_{\nu\pm1}(\tau)$ is  obtained from $D_{\nu}(\tau)$ in (6-7) by
substitution $m^2\to m^2\mp2ieE$. The arbitrary coefficients $c_1, c_2,%
c_{\pm}\equiv{}_+c_{\pm}$ determine polarization of vector boson. They are not
independent due to the constraint (2):
\begin{equation}
c_1p_1+c_2p_2+\sqrt{2eE}e^{\frac{i\pi}{4}}[(1+\nu){}_+c_+-{}_+c_-]=0.
\end{equation}                                              
For negative-frequency in-solution (for scalar boson) instead of (6) we have
\begin{equation}
{}_-\psi\propto[D_{\nu}(\tau)]^*e^{i\vec p\cdot\vec x}.               
\end{equation}
Star means complex conjugation. Similarly to (8) the parabolic cylinder
functions entering in ${}_-\psi^{\pm}$ are obtained from
$[D_{\nu}(\tau)]^*$ in (11) by substitutions $m^2\to m^2{\mp}2ieE$. So
$$
{}_-\psi=\begin{bmatrix}
c_+D_{\nu^*-1}(\tau^*)+c_-D_{\nu^*+1}(\tau^*)\\
\hspace{0.15cm}c_1D_{\nu^*}(\tau^*)\\
\hspace{0.15cm}c_2D_{\nu^*}(\tau^*)\\
c_+D_{\nu^*-1}(\tau^*)-c_-D_{\nu^*+1}(\tau^*)
\end{bmatrix}e^{i\vec p\cdot\vec x}.                      \eqno(12)
$$
(In (12) $c_{\pm}={}_{-}c_{\pm}$ and similarly in other cases)
The constraint takes the form
$$
c_1p_1+c_2p_2+\sqrt{2eE}e^{-\frac{i\pi}{4}}[{}_-c_++\nu{}_- c_-]=0.                                                  \eqno(13)
$$
 Nothing prevents us to assume that $c_1, c_2$ in (12) are the same as in (8).

The negative-frequency out-solution is obtained from positive frequency
in-solution by changing sign of $\tau$ in parabolic cylinder functions in
(8):
$$
{}^-\psi=\begin{bmatrix}
c_+D_{\nu+1}(-\tau)+c_-D_{\nu-1}(-\tau)\\
\hspace{0.15cm}c_1D_{\nu}(-\tau)\\
\hspace{0.15cm}c_2D_{\nu}(-\tau)\\
c_+D_{\nu+1}(-\tau)-c_-D_{\nu-1}(-\tau)
\end{bmatrix}e^{i\vec p\cdot\vec x},\quad
{}^-c_{\pm}=-{}_+c_{\pm}, \eqno(14)
$$
see (112a).
The constraint has the form
$$
c_1p_1+c_2p_2+\sqrt{2eE}e^{i\frac{\pi}{4}}[{}^-c_--(1+\nu){}^-c_+]=0. \eqno(15)
$$

Similarly we find the positive frequency out-solution from ${}_-\psi$ in (12)
by changing sign of $\tau^*$
$$
{}^+\psi=\begin{bmatrix}
c_+D_{\nu^*-1}(-\tau^*)+c_-D_{\nu^*+1}(-\tau^*)\\
\hspace{0.15cm}c_1D_{\nu^*}(-\tau^*)\\
\hspace{0.15cm}c_2D_{\nu^*}(-\tau^*)\\
c_+D_{\nu^*-1}(-\tau^*)-c_-D_{\nu^*+1}(-\tau^*)
\end{bmatrix}e^{i\vec p\cdot\vec x}.                   \eqno(16)
$$
The corresponding constraint is
$$
c_1p_1+c_2p_2-\sqrt{2eE}e^{-i\frac{\pi}{4}}[\nu{}^+c_-+{}^+c_+]=0.  \eqno(17)
$$

Now for scalar boson the in- and out-solutions are related by [6]
\setcounter{equation}{17}
\begin{gather}
{}_+\psi_n=c_{1n}{}^+\psi_n+c_{2n}{}^-\psi_n,\\        \notag
{}_-\psi_n=c_{2n}^*{}^+\psi_n+c_{1n}^*{} ^-\psi_n,\\   \notag
c_{1n}=\frac{\sqrt{2\pi}}{\Gamma((1-i\lambda)/2)}\exp[-\frac{\pi}{4}(\lambda-i)],
\quad c_{2n}=\exp[-\frac{\pi}{2}(\lambda+i)],\\            \notag
 \vert c_{1n}\vert^2- \vert c_{2n}\vert^2=1.                             
\end{gather}
Subscript $n$ means a set of quantum numbers. Here $n=\vec p$.
  By straightforward calculation
similar to scalar case
we find that (18) hold also for vector boson and that
\begin{gather}
{}^+c_-=\frac{i}{\nu}{}_+c_-=-{}_-c_-=-\frac{i}{\nu}{}^-c_-,\\ \notag
{}^+c_+=-i(1+\nu){}_+c_+=-{}_-c_+=i(1+\nu){}^-c_+.                                                               
\end{gather}
These relations guarantee that all  wave functions ${}_{\pm}\psi,
{}^{\pm}\psi$ are normalized in the same manner and that any constraint
can be obtained from any other using (19).

As seen from (18) in constant electric field the Bogoliubov coefficients
$c_{1n},\;c_{2n}$ does not
depend on boson polarization. Thus the imaginary part of the Lagrange function
is simply $3{\rm ImL}_{spin0}$ in agreement with [3,4].

\section{Vector boson in constant electromagnetic field}
Now we add to constant electric field the collinear constant magnetic field.
Assuming $A_2=Hx_1$ we have for $\psi_1$ and $\psi_2$ from (1)
\begin{equation}
(-D^2+m^2)\psi_1-2ieH\psi_2=0.\quad (-D^2+m^2)\psi_2+2ieH\psi_1=0.  
\end{equation}
 Introducing
\begin{equation}
\tilde\psi_1=\psi_1-i\psi_2,\quad\tilde\psi_2=\psi_1+i\psi_2,\quad
\psi_1=\frac12(\tilde\psi_1+\tilde\psi_2),                             
\quad\psi_2=\frac{i}{2}(\tilde\psi_1-\tilde\psi_2),
\end{equation}
we rewrite (20) as
\begin{equation}
(-D^2+m^2+2eH)\tilde\psi_1=0,\quad (-D^2+m^2-2eH)\tilde\psi_2=0,            
\end{equation}
i.e. $\tilde\psi_{1,2}$ are obtained from scalar boson wave function by
substitutions $m^2\to m^2\pm2eH$. So we may write
\begin{equation}
\tilde\psi_1\propto2c_1D_{n-1}(\zeta),\quad\tilde\psi_2\propto2c_2D_{n+1}(\zeta)
\quad\zeta=\sqrt{2eH}(x_1+\frac{p_2}{eH}).                                         
\end{equation}
 Thus we have instead of (8)
$$
{}_+\psi_{p_2 p_3 n}=\begin{bmatrix}
[c_+D_{\nu+1}(\tau)+c_-D_{\nu-1}(\tau)]D_{n}(\zeta)\\

[c_1D_{n-1}(\zeta)+c_2D_{n+1}(\zeta)]D_{\nu}(\tau)\\
i[c_1D_{n-1}(\zeta)-c_2D_{n+1}(\zeta)]D_{\nu}(\tau)\\

[c_+D_{\nu+1}(\tau)-c_-D_{\nu-1}(\tau)]D_{n}(\zeta)
\end{bmatrix}
e^{i(p_2x_2+p_3x_3)},    \eqno(24)
$$
and similarly for other $\psi$. Here
$$
\nu=\frac{i\lambda}2-\frac12,\quad
 \lambda=\frac{m^2+eH(2n+1)}{eE},
  \eqno(25)
$$

The constraints can be obtained from previous ones by substitution
$$
c_1p_1+c_2p_2\to-i\sqrt{2eH}[(1+n)c_2-c_1].     \eqno(26)
$$
We note here that $D_{\mu}\psi^{\mu}$ is proportional to scalar wave function
$$
D_n(\zeta)D_{\nu}(\tau)\exp[i(p_2x_2+p_3x_3)]
$$ (which is dropped in
the expressions like (10) with modification (26), or in (116)). The equations (67)
 and (98) were used to obtain the constraints. It follows from derivation that
the presence of $c_1$ in the r.h.s. of (26) is due to the assumption that
$D_{n-1}(\zeta)$ in (24) is not zero, i.e. $n\ge1$.

Using (24) and (26), we can build three polarization states $\psi(i,x),
 i=1, 2, 3$, see Sec. 7. For these states the minimal values of $n$ in (25)
are $-1, 0, 1$ correspondingly. Thus the Bogoliubov coefficients depend
 on all four quantum numbers $(n=p_2,p_3,n,i)$ through minimal $n$.

Taking into account that $2{\rm ImL}=\sum_n\ln(1+\vert c_{2n}\vert^2)$
 it is easy to show that in agreement with [4]
$$
{\rm Im 2L}_{spin1}=2\times{\rm 3Im L}_{spin0}
+\{\ln[1+e^{-\pi\frac{m^2-eH}{eE}}]-\ln[1+e^{-\pi\frac{m^2+eH}{eE}}]\}%
\frac{\alpha}{\pi}EHVT.    \eqno(27)
$$
The factors outside braces give the statistical weight of "correcting"
states, see eqs. (3.6),
(3.7) in [6].

The Bogoliubov coefficients permit to find the transition
  probability from any initial to  any final
state (with any occupation numbers) [6]. For example, if the initial state is
vacuum, we have for the cell with set of quantum numbers $n=p_2,p_3,n,i$
$$
\vert c_{1n}\vert ^{-2}\{1+w_n+w_n^2+w_n^3+\cdots\}=1,\quad w_n=%
\frac{\vert c_{2n}\vert^2}{\vert c_{1n}\vert^2}.                    \eqno(28)
$$
Term $\vert c_{1n}\vert^{-2}w_n^k$ gives the probability for production of
$k$ pairs, $k=0, 1, 2, \cdots$. The events in cells with different
quantum numbers are independent.

 \section{ Propagator of free vector boson}

We may take the wave functions of a free vector boson with momentum
$p^{\mu}=(p^0,0,0,p^3)$ in the form
$$
\psi^{\mu}(i,x)=\frac{u^{\mu}(i)}{\sqrt{2\vert p^0\vert}}e^{ip\cdot x},
\quad \eta^{\mu\nu}={\rm diag}(-1,1,1,1) ,\quad \mu=0,1,2,3, \eqno(29)
$$
$$
u(1)=\begin{bmatrix}
0\\
1\\
0\\
0
\end{bmatrix},\quad u(2)=\begin{bmatrix}
0\\
0\\
1\\
0
\end{bmatrix},\quad u(3)=\frac1m\begin{bmatrix}
p_3\\
0\\
0\\
p^0
\end{bmatrix}.
$$
These solutions satisfy the wave equation (1) and constraint (2) with
switched off external field.
Summing $\psi^{\mu}(i,x)\psi^{\nu*}(i,x')$ over polarization, we find
$$
\sum_{i=1}^{3}\psi^{\mu}(i,x)\psi^{\nu*}(i,x')=\frac1{2\vert p^0\vert}
\begin{bmatrix}
\frac{p_3^2}{m^2}& 0& 0& \frac{p_3p^0}{m^2}\\
0& 1& 0& 0\\
0& 0& 1& 0\\
\frac{p_3p^0}{m^2}& 0& 0& \frac{(p^{0})^2}{m^2}
\end{bmatrix}e^{i(p,x-x')} \eqno(30)
$$

If instead of linear polarization states (29) we use helicity states
 (cf. \S16 in
[8]), we get the same result (30). In general we have to replace the
matrix in the r.h.s. of (30) by $\eta^{\mu\nu}+p^{\mu}p^{\nu}/m^2$. Only the
presence of this matrix in the expression for propagator (similar to (51))
differs this case from scalar particle case. For this reason the vector
 boson propagator can be obtained from scalar one
$$
G_{spin0}(x,x')=\frac1{(2\pi)^4}\int d^4p\frac{e^{i(p,x-x')}}{p^2+m^2-
i\varepsilon}=
\frac1{(4\pi)^2}\int\limits_0^\infty\frac{ds}{s^2}e^{-ism^2+\frac{i(x-x')^2}
{4s}},\eqno(31)
$$
considered as a unit matrix over discrete indices, by acting
on it with the differential operator:
$$
  G^{\mu\nu}(x,x')=\left(\eta^{\mu\nu}-\frac1{m^2}\frac{\partial^2}%
{\partial x_{\mu}\partial x_{\nu}}\right)G_{spin0}(x,x'). \eqno(32)
$$

As the scalar boson propagator satisfies the equation
$$
(-\partial_{\mu}\partial^{\mu}+m^2)G_{spin0}(x,x')=\delta^4(x-x'), \eqno(33)
$$
we have for vector boson
$$
(-\partial_{\sigma}\partial^{\sigma}+m^2)G^{\mu\nu}(x,x')=(\eta^{\mu\nu}-
\frac1{m^2}\frac{\partial^2}{\partial x_{\mu}\partial x_{\nu}})\delta^4(x-x'),
\eqno(34)
$$
i.e. on the r.h.s. stands not simply $\delta^4(x-x')$. The complication
is due to the existence of constraint. This circumstance
prevent us from using the well-known methods of constructing propagators
of scalar and spinor particles in an external field [9, 10]. An elegant
way to circumvent this difficulty is given by Vanyashin and Terentyev [3].

\section{Vector boson propagator in a constant magnetic field}

To write down the propagator we need the complete set of orthonormalized
solutions. The orthonormalization is performed with the help of expression
for vector current [5]
$$
J_{\mu}=-i\{\psi^{\nu*}(D_{\mu}\psi_{\nu}-D_{\nu}\psi_{\mu})-(D^*_{\mu}\psi^*
_{\nu}-D^*_{\nu}\psi^*_{\mu})\psi^{\nu}\},\quad D_{\mu}=\partial_{\mu}+
ieA_{\mu}.\eqno(35)
$$

 By the way our expression for $D_{\mu}$ in (35) coincide with that in [5];
although our $\eta_{\mu\nu}$ differs in sign, we also substitute
$e\to-e$, using $e=|e|$ and assuming by analogy with electron that
$W^-$ is particle.

In space  without field the expression for $J_{\mu}$ up to divergence terms
can be written in form similar  to scalar case, see \S 15 in [8]. It is
remarkable that with constraint (2) the same is true in the presence of field.
Indeed,
$$
-\psi^{\nu*}D_{\nu}\psi_{\mu}=-\partial_{\nu}(\psi^{\nu*}\psi_{\mu})+
\psi_{\mu}D^*_{\nu}\psi^{\nu*}.\eqno(36)
$$
But the last term on the r.h.s. is zero due to (2) for boson with
$g=2$. Similarly
$$
(D^*_{\nu}\psi^*_{\mu})\psi^{\nu}=\partial_{\nu}(\psi^{\nu}\psi^*_{\mu})
-\psi^*_{\mu}D_{\nu}\psi^{\nu}=   \partial_{\nu}(\psi^{\nu}\psi^*_{\mu}).
\eqno(37)
$$
So
$$
J_{\mu}=-i\{\psi^{\nu*}D_{\mu}\psi_{\nu}-(D^*_{\mu}\psi^*_{\nu})\psi^{\nu}
-\partial_{\nu}
[\psi^{\nu*}\psi_{\mu}-\psi^{\nu}\psi^*_{\mu}]\}.\eqno(38)
$$
 To normalize wave functions we need only $J_0$. Straightforward calculations
show that the divergence terms does not contribute to $J_0$ for considered fields.  Then
$$
J^0=-J_0=i\{\psi^{\nu*}D_0\psi_{\nu}-(D^*_0\psi^*_{\nu})\psi^{\nu}\}.\eqno(39)
$$
Correspondingly we use for orthonormalization the expression
$$
J^0(\psi',\psi)=i\{\psi'{}^{\nu*}D_0\psi_{\nu}-(D^*_0\psi'{}^*_{\nu})
\psi^{\nu}\}
\eqno(40)
$$

 For our vector-potentials $A_0(x)=0$. Then $D_0=\frac{\partial}{\partial t}
\quad$
 and
$$
J^0(\psi',\psi)=i\{\psi'{}^*_{k}\buildrel\leftrightarrow\over\partial_t\psi_k-
\psi'{}^{0*}                      \buildrel\leftrightarrow\over\partial_t\psi^0\}
\eqno(41)
$$
The sum over $k$ runs from 1 to 3.

The positive-frequency solution of the wave equation (1) with
$A_{\mu}(x)=\delta_{\mu2}Hx_1$ has the form
$$
\psi^{\mu}_{p_2,p_3,n}=\begin{bmatrix}
c^0D_n(\zeta)\\
c_1D_{n-1}(\zeta)+c_2D_{n+1}(\zeta)\\
i[c_1D_{n-1}(\zeta)-c_2D_{n+1}(\zeta)]\\
c_3D_n(\zeta)
\end{bmatrix}e^{i(p_2x_2+p_3x_3-p^0t)}.\eqno(42)
$$
The elements of this column correspond to $\mu=0,1,2,3$,
$$
 \zeta=\sqrt{2eH}(x_1+\frac{p_2}{eH}),\quad
p^0=\sqrt{m^2+p_3^2+eH(2n+1)}.
$$
  The coefficients $c$, determining the boson polarization, satisfy the
constraint
$$
-ip^0c^0+ip_3c_3+\sqrt{2eH}[(n+1)c_2-c_1]=0. \eqno(43)
$$

For states with polarizations $c'$ and $c$ we get from (41) and (42)
$$
J^0(\psi',\psi)=2p^0\{2c'{}^*_1c_1D_{n-1}^2(\zeta)+2c'{}^*_2c_2D_{n+1}^2(\zeta)
+(c'{}^*_3c_3-c'{}^{0*}c^0)D_n^2(\zeta)\},\eqno(44)
$$
Integrating over $x_1$, we find
$$
\int\limits_{-\infty}^{\infty}dx_1J^0(\psi',\psi)=2p^0n!\sqrt{\frac{\pi}{eH}}
\{\frac2nc'{}^*_1c_1+2(n+1)c'{}^*_2c_2+c'{}^*_3c_3-c'{}^{0*}c^0\},\eqno(45)
$$
$$
\int\limits_{-\infty}^{\infty}dx_1D_n^2(\zeta)=n!\sqrt\frac{\pi}{eH}.
$$

Using orthonormalization conditions
$$
\int dx_1J^0({}_{\pm}\psi(i,x),{}_{\pm}\psi(j,x))={}\pm\delta_{ij},\quad
i,j=1,2,3,  \eqno(46)
$$
and constraint (43), we find the following positive-frequency polarization
states
$$
\psi^{\mu}(1,x)=N(1)\begin{bmatrix}
(n+1)\sqrt{2eH}p^0D_n(\zeta)\\
im_{\perp}^2D_{n+1}(\zeta)\\
m^2_{\perp}D_{n+1}(\zeta)\\
(n+1)\sqrt{2eH}p_3D_n(\zeta)
\end{bmatrix}e^{i(p_2x_2+p_3x_3-p^0t)}.\eqno(47)
$$
Here
$$
\mu=0,1,2,3,\quad m^2_{\perp}=m^2+eH(2n+1),\quad p^0=\sqrt{m^2+p^2_3+eH(2n+1)},
$$
$$
N(1)=n_1N_0,\quad N_0=\left(\frac{eH}{\pi}\right)^{\frac14}\frac1{\sqrt{2|p^0|n!}},\quad
n_1=\frac1{\sqrt{2(n+1)m^2_{\perp}(m^2+eHn)}}.\eqno(48)
$$
$$
\psi(2,x)=n_2N_0\begin{bmatrix}
p_3D_n(\zeta)\\
0\\
0\\
p^0D_n(\zeta)\end{bmatrix}e^{i(p_2x_2+p_3x_3-p^0t)},\quad n_2=
\frac1{\sqrt{m^2_{\perp}}},\eqno(49)
$$

$$
\psi(3,x)=n_3N_0\begin{bmatrix}
\sqrt{2eH}p^0D_n(\zeta)\\
i[-(m^2+eHn)D_{n-1}(\zeta)+eHD_{n+1}(\zeta)]\\
(m^2+eHn)D_{n-1}(\zeta)+eHD_{n+1}(\zeta)\\
\sqrt{2eH}p_3D_n(\zeta)\end{bmatrix}e^{i(p_2x_2+p_3x_3-p^0t)},
$$
$$
n_3=\sqrt{\frac n{2m^2(m^2+eHn)}}.\eqno(50)
$$
 We detach from normalization factors $N(i)$ the normalization factor
$N_0$ of scalar wave function, because we shall concentrate our attention
on the differences from scalar case. We note also that
 $N(3)\propto\Gamma^{-\frac12}(n)$ and is zero for $n=0$. So for state
$\psi(3,x)$ only $n=1, 2, 3\cdots$ are possible.  The same follows from the
fact that the constraint (43) cannot be satisfied because $c_1$ is absent in it
for $n=0$.

Now we are in a position to build up vector boson propagator. We start from the
 expression (which is a special case of more general result derived in Sec. 6,
see (80-81))
$$
G^{\mu\nu}(x,x')=i\int\limits_{-\infty}^{\infty}\frac{dp_2}{2\pi}
\int\limits_{-\infty}
^{\infty}\frac{dp_3}{2\pi}\sum_{n=-1}^{\infty}\sum_{i=1}^{3}\left\{
\begin{array}{cc}
{}_+\psi^{\mu}(i,x){}_+\psi^{*\nu}(i,x'),\quad t>t'\\
{}_-\psi^{\mu}(i,x){}_-\psi^{*\nu}(i,x'), \quad t<t'
\end{array}\right.\eqno(51)$$

We denote in the following the previous quantity $p^0$ as $E_n$ and use
 the relations
$$
-\frac1{2\pi i}\int\limits_{-\infty}^{\infty}dp^0\frac{e^{-ip^0(t-t')}}
{(p^0-E_n+i\epsilon)(p^0+E_n-i\epsilon)}=\frac1{2E_n}\left\{\begin{array}{cc}
e^{-iE_n(t-t')},\quad t>t'\\
e^{iE_n(t-t')},\quad t<t'\end{array}\right.,\eqno(52)
$$
$$
\frac1{i(E_n^2-(p^0)^2)}=\int\limits_0^{\infty}dse^{-is(E_n^2-p_0^2)}
$$
to rewrite (51) in the form $(p^0=-p_0)$
$$
G^{\mu\nu}(x,x')=i\sqrt{\frac{eH}{\pi}}\sum_{n=-1}^{\infty}\int\limits_{-\infty}
^{\infty}\frac{dp_2}{2\pi}\int\limits_{-\infty}^{\infty}\frac{dp_3}{2\pi}
\int\limits_{-\infty}^{\infty}\frac{dp^0}{2\pi}\int\limits_0^{\infty}ds
a^{\mu\nu}(x,x')\frac1{n!}\times
$$
$$
e^{-is(m^2_{\perp}+p^2_3-p_0^2)+i[p_2(x_2-x'_2)+
p_3(x_3-x'_3)-p^0(t-t')]},\quad m_{\perp}^2=m^2+eH(2n+1).\eqno(53)
$$
We note that the lower line on r.h.s. of (52) is obtained from the upper
line by substitution $t\leftrightarrow t'$, which  does not change anything,
as the r.h.s. may be written in the form $(2E_n)^{-1}\exp[-iE_n|t-t'|]$.
The explicitly symmetric in $t,t'$ form of the l.h.s. is
$$
\int\limits_0^{\infty}dse^{-isE^2_n}\int\limits_{-\infty}^{\infty}\frac
{dp^0}{2\pi}e^{isp_0^2-ip^0(t-t')}=\frac{e^{\frac{i\pi}{4}}}{2\sqrt\pi}
\int\limits_0^{\infty}\frac{ds}{\sqrt s}\exp[-isE_n^2-i\frac{(t-t')^2}{4s}].
\eqno(54)
$$

We first get the propagator for scalar particle in proper time representation
 [10]. We
 substitute  in (53) $a^{\mu\nu}(x,x')$ by $D_n(\zeta)D_n(\zeta')$. Then
using the formula
$$
D_n(\zeta)=\sqrt{\frac2{\pi}}e^{\frac{\zeta^2}{4}}\int\limits_0^{\infty}
dyy^ne^{\frac{-y^2}{2}}\cos(\zeta y-\frac{n\pi}{2}),\eqno(55)
$$
we find
$$
\sum_{n=0}^{\infty}\frac{D_n(\zeta)D_n(\zeta')}{n!}e^{-i\tau(2n+1)}=
(2\sin2\tau)^{-\frac12}\exp[-i\frac{\pi}{4}+i\frac{(\zeta-\zeta')^2}{8\tan\tau}-
i\frac{(\zeta+\zeta')^2}{8\cot\tau}], $$
$$
\tau=eHs,\quad \zeta'=\sqrt{2eH}(x'_1+\frac{p_2}{eH}).\eqno(56)
$$
Subsequent integration over $p_2$ gives
$$
   \int\limits_{-\infty}^\infty\frac{dp_2}{2\pi}\sum_{n=0}^{\infty}\frac
{D_n(\zeta)D_n(\zeta')}{n!}e^{-i\tau(2n+1)+ip_2z_2}=
$$
$$
-i\sqrt{\frac{eH}{\pi}}
(4\sin\tau)^{-1}\exp[-i\frac{eHz_2(x_1+x'_1)}{2}+i\frac{eH(z_1^2+z_2^2)}
{4\tan\tau}],\quad z_{\mu}=x_{\mu}-x'_{\mu}. \eqno(57)
$$
Using also
$$
\int\limits_{-\infty}^{\infty}\frac{dp_3}{2\pi}\int\limits_{-\infty}^{\infty}
\frac{dp^0}{2\pi}\exp[i(p_3z_3-p^0z^0)-is(p_3^2-p_0^2)]=\frac1{4\pi s}
\exp[i\frac{z_3^2-z_0^2}{4s}], \eqno(58)
$$
we find [9, 10, 6, 11].
$$
G_{spin0}(x,x')=\frac{eH}{(4\pi)^2}\int\limits
_0^{\infty}\frac{ds}{s\sin eHs}\times
$$
$$
\exp[-i\frac{eHz_2(x_1+x'_1)}{2}]\exp[-ism^2+i\frac{z_3^2-z_0^2}{4s}+
i\frac{(z_1^2+z_2^2)eH}{4\tan eHs}].\eqno(59)
$$

 We shall show now how to obtain $a^{\mu\nu}(x,x')$ in (53) and how to turn it
into differential matrix, which, when inserted in the integrand in (59), will
give the propagator of vector boson. As a preliminary we write down two
formulae directly related to (56):
$$
\sum_{n=-1}^{\infty}\frac{D_{n+1}(\zeta)D_{n+1}(\zeta')}{(n+1)!}e^{-i\tau(2n+1)}
=e^{2i\tau}\sum_{n=0}^{\infty}\frac{D_n(\zeta)D_n(\zeta')}{n!}e^{-i\tau(2n+1)},
\eqno(60)
$$
$$
\sum_{n=1}^{\infty}\frac{D_{n-1}(\zeta)D_{n-1}(\zeta')}{(n-1)!}e^{-i\tau(2n+1)}=
e^{-2i\tau}\sum_{n=0}^{\infty}\frac{D_n(\zeta)D_n(\zeta')}{n!}e^{-i\tau(2n+1)}.
\eqno(61)
$$
We see that the expressions in (60) and (61) only by factors $e^{2i\tau}$ and
$e^{-2i\tau}$  differ from scalar case.

Now we go back to $a^{\mu\nu}(x,x')$. As seen from (51) and (53)
$$
a^{\mu\nu}(x,x')\propto\sum_{i=1}^3\psi^{\mu}(i,x)\psi^{\nu*}(i,x').\eqno(62)
$$
For example, we consider $a^{11}(x,x')$. We see from (49) that $\psi^1(2,x)=0$,
 i.e. the term with $i=2$ does not contribute to $a^{11}(x,x')$. According
to (47) the contribution from term with $i=1$ is
$$
n_1^2m_{\perp}^4D_{n+1}(\zeta)D_{n+1}(\zeta'),\quad n_1^2=\frac1
{2(n+1)m_{\perp}^2(m^2+eHn)}.\eqno(63)
$$
Term with $i=3$ gives
$$
n_3^2[-(m^2+eHn)D_{n-1}(\zeta)+eHD_{n+1}(\zeta)][-(m^2+eHn)D_{n-1}(\zeta')+
eHD_{n+1}(\zeta')],
$$
$$
 n_3^2=\frac{n}{2m^2(m^2+eHn)}.\eqno(64)
$$
Now we have $a^{11}(x,x')$ as a sum of (63) and (64):
$$
a^{11}(x,x')=\frac1{2(m^2+eHn)}\left(\frac{m_{\perp}^2}{n+1}+\frac{(eH)^2n}
{m^2}\right)D_{n+1}(\zeta)D_{n+1}(\zeta')+
$$
$$
\frac{n(m^2+eHn)}{2m^2}D_{n-1}(\zeta)D_{n-1}(\zeta')-\frac{eHn}{2m^2}%
[D_{n+1}(\zeta)D_{n-1}(\zeta')+D_{n-1}(\zeta)D_{n+1}(\zeta'].\eqno(65)
$$
 Next we note that
$$
\frac1{m^2+eHn}\left(\frac{m_{\perp}^2}{n+1}+\frac{(eH)^2n}{m^2}\right)
=\frac1{n+1}+\frac{eH}{m^2},\eqno(66)
$$
i.e. the undesirable factor $\frac1{m^2+eHn}$, contained in $n_1^2$ and
$n_3^2$ in (63) and (64), disappears  in the sum (65).

Further we use the relations, see eqs.(8.2.15-16) in [7]
$$
\left(\frac{d}{d\zeta}+\frac{\zeta}{2}\right)D_n(\zeta)=nD_{n-1}(\zeta),\quad
\left(\frac{d}{d\zeta}-\frac{\zeta}{2}\right)D_n(\zeta)=-D_{n+1}(\zeta).\eqno(67)
$$
We write down also the sum and difference of these expressions:
$$
2\frac{d}{d\zeta}D_n(\zeta)=nD_{n-1}(\zeta)-D_{n+1}(\zeta),\quad
\zeta D_n(\zeta)=nD_{n-1}(\zeta)+D_{n+1}(\zeta).\eqno(68)
$$

Now it is easy to verify that
$$
a^{11}(x,x')=\frac{D_{n+1}(\zeta)D_{n+1}(\zeta')}{2(n+1)}+\frac n2D_{n-1}(\zeta)
D_{n-1}(\zeta')+\frac{2eH}{m^2}\frac{\partial^2}{\partial\zeta\partial\zeta'}
D_n(\zeta)D_n(\zeta').\eqno(69)
$$
The first term on the r.h.s. of (69) works in eq. (60), the second term
is used in (61); the necessary factor $n!$ comes from $N_0$, see (48). The
third term can be written as
$$
\frac1{m^2}\frac{\partial^2}{\partial x_1\partial x'_1}D_n(\zeta)D_n(\zeta').
\eqno(70)
$$
In a similar manner we find the other $a^{\mu\nu}(x,x')=a^{\nu\mu*}(x',x)$.
It is easy to verify that the differential operator $A^{\mu\nu}(x,x')$,
corresponding to $a^{\mu\nu}(x,x')$, has the form
$$
A^{\mu\nu}=B^{\mu\nu}+C^{\mu\nu},\quad C^{\mu\nu}=\frac1{m^2}\Pi^{\mu}(x)%
\Pi^{\nu*}(x'),
$$
$$
 \Pi_{\mu}(x)=
-i\frac{\partial}{\partial x^{\mu}}+eA_{\mu}(x),\quad \Pi^*_{\mu}(x')=
 i\frac{\partial}{\partial x'{}^{\mu}}+eA_{\mu}(x')  .\eqno(71)
$$
In our case
$$
A_{\mu}(x)=\delta_{\mu2}Hx_1,
\quad \Pi^0(x)=i\frac{\partial}{\partial t},
\quad \Pi^{0*}(x')=-i\frac{\partial}{\partial t'}.\eqno(72)
$$
The nonzero $B^{\mu\nu}$ are
$$
B^{11}=B^{22}=\cos\tau,\quad B^{21}=-B^{12}=\sin\tau,\quad B^{33}=-B^{00}=1.
\eqno(73)
$$
The difference of $B^{\mu\nu}$ from $\eta^{\mu\nu}$ is due to the interaction of
 boson magnetic moment with magnetic field. We may say that $B^{\mu\nu}$ with
$\mu,\nu=1,2$ are "dressed" by magnetic field.

Thus
$$
G^{\mu\nu}(x,x')=\frac{eH}{(4\pi)^2}\int\limits_0^%
{\infty}\frac{ds}{s\sin eHs}e^{-ism^2}%
A^{\mu\nu}\times
$$
$$
\exp[-\frac{ieHz_2(x_1+x'_1)}{2}]\exp[\frac{i(z_3^2-z_0^2)}{4s}+
\frac i4(z_1^2+z_2^2)eH\cot eHs],\quad z_{\mu}=x_{\mu}-x'_{\mu}.\eqno(74)
$$
It is somewhat surprising that this representation does not coincide with
Vanyashin- Terentyev representation [3] with switched off electric field.
 Possibly these are two different
representations of one and the same propagator and it would be interesting
 to verify this supposition.

\section{Propagator of vector boson in constant electric field}

First, we shall give the generalization of (51) for the case, when external
field can create pairs [12, 6]. To this end  we write
$$
G(x,x')_{abs}=i<0_{out}|T(\Psi(x)\Psi^{\dagger}(x'))|0_{in}>=
<0_{out}|0_{in}>G(x,x'), \eqno(75)
$$
where $T$ is chronological ordering operator,
$$
\Psi(x)=\sum_n[a_{n\,out}\;{}^+\psi_n(x)+b_{n\,out}^+\;{}^-\psi_n(x)],
$$
$$
\Psi^{\dagger}(x)=\sum_n[a_{n\,in}^\dagger\;{}_+\psi_n^*(x)+b_{n\,in}\;
{}_-\psi_n^*
(x)]\eqno(76)
$$
As usual, $a_n$ and $b_n$ are operators of destruction of particle and
antiparticle in a state with quantum numbers $n$:
$$
\Psi^{\dagger}(x')|0_{in}>=\sum_k{}_+\psi_k^*(x')a_{k\,in}^{\dagger}|0_{in}>,
\: <0_{out}|\Psi(x)=<0_{out}|\sum_na_{n\,out}\;{}^+\psi_n(x).\eqno(77)
$$
For $t>t'$ from (75), (77) we have
$$
G(x,x')_{abs}=i\sum_{n,k}{}^+\psi_n(x)\;{}_+\psi_k^*(x')<0_{out}|a_{n\,out}
a_{k\,in}^{\dagger}|0_{in}>,\:t>t'.\eqno(78)
$$
 The Bogoliubov
transformations in our case have the form (cf. eq. (18))[6]
$$
a_{n\,out}^{\dagger}=c_{1\,n}^*a_{n\,in}^{\dagger}+c_{2\,n}b_{n\,in},\quad
b_{n\,out}=          c_{2\,n}^*a_{n\,in}^{\dagger}+c_{1\,n}b_{n\,in}.\eqno(79)
$$
From the first eq. in (79) it follows $a_{k\,out}^{\dagger}|0_{in}>=
c_{1\,k}^*a_{k\,in}^{\dagger}|0_{in}>$. We substitute
 $a_{k\,in}^{\dagger}|0_{in}>$  from here into (78) and use the commutation
relation $[a_{k\,out},a_{n\,out}^{\dagger}]=\delta_{kn}.$ Then we obtain
$$
G(x,x')_{abs}=<0_{out}|0_{in}>i\sum_n{}^+\psi_n(x)\;{}_+\psi_n^*(x')\frac
1{c_{1\,n}^*},\quad t>t'. \eqno(80)
$$
Similarly for $t<t'$ we find
$$
G(x,x')_{abs}=<0_{out}|0_{in}>i\sum_n{}_-\psi_n(x)\;{}^-\psi_n^*(x')\frac%
1{c_{1\,n}^*},\quad t<t'. \eqno(81)
$$
If an external field does not create pairs, the obtained expressions go into
(51).

The transition current (41) in terms of states ${}_+\psi',{}_+\psi$ in (8)
takes the form
$$
J^0({}_+\psi',{}_+\psi)=\sqrt{2eE}e^{\frac{\pi\lambda}4}[c'{}^*_1c_1+c'{}^*_2
c_2+2i({}_+c'{}^*_-\;{}_+c_+-{}_+c'{}^*_+\;{}_+c_-)].\eqno(82)
$$
Here use has been made of eq. (8.2.11) in [7] (and its complex conjugate):
$$
D_{\nu^*+1}(\tau^*)\buildrel\leftrightarrow\over{\frac d{dt}}D_{\nu-1}(\tau)=
\sqrt{2eE}\exp[\frac{\pi\lambda}{4}]= \eqno(83)
$$
$$
-D_{\nu^*-1}(\tau^*)
                   \buildrel\leftrightarrow\over{\frac d{dt}}D_{\nu+1}(\tau)=
D_{\nu^*}(\tau^*)i \buildrel\leftrightarrow\over{\frac d{dt}}D_{\nu}(\tau).
\eqno(84)
$$

The constraint is given in (10). Using (82), (8) and (10) we find the following
 ${}_+\psi$ polarization states:
$$
{}_+\psi(1,x)=N(1)
\begin{bmatrix}
p_2\sqrt{\frac{eE}2}e^{\frac{i\pi}4}[D_{\nu+1}(\tau)-\nu D_{\nu-1}(\tau)]\\
                                 0\\
m_{\perp}^2D_{\nu}(\tau)\\
p_2\sqrt{\frac{eE}2}e^{\frac{i\pi}4}[D_{\nu+1}(\tau)+\nu D_{\nu-1}(\tau)]
\end{bmatrix}e^{i\vec p\cdot\vec x},\quad N(i)=n_iN_0,
$$
$$
N(1)=n_1N_0,\quad n_1=\sqrt{\frac1{m_{\perp}^2(m^2+p_1^2)}},\quad
N_0=(2eE)^{-\frac14}e^{-\frac{\pi\lambda}8},\eqno(85)
$$
$$
{}_+\psi(2,x)=N(2)\begin{bmatrix}
D_{\nu+1}(\tau)+(1+\nu)D_{\nu-1}(\tau)\\
0\\
0\\
D_{\nu+1}(\tau)-(1+\nu)D_{\nu-1}(\tau)
\end{bmatrix}e^{i\vec p\cdot\vec x},
$$
$$
n_2=\sqrt{\frac{eE}{2m_{\perp}^2}},\quad m_{\perp}^2=m^2+p_1^2+p_2^2,\eqno(86)
$$
$$
{}_+\psi(3,x)=N(3)\begin{bmatrix}
p_1\sqrt{\frac{eE}2}e^{\frac{i\pi}4}[D_{\nu+1}(\tau)-\nu D_{\nu-1}(\tau)]\\
(m^2+p_1^2)D_{\nu}(\tau)\\
p_1p_2D_{\nu}(\tau)\\
p_1\sqrt{\frac{eE}2}e^{\frac{i\pi}4}[D_{\nu+1}(\tau)+\nu D_{\nu-1}(\tau)]
\end{bmatrix}e^{i\vec p\cdot\vec x},
$$
$$
n_3=\frac1{\sqrt{m^2(m^2+p_1^2)}},
\eqno(87)
$$

The ${}^+\psi$ polarization states can be obtained from these ones with the
 help of eqs. (19) (see also (16)):
$$
{}^+\psi(1,x)=N(1)\begin{bmatrix}
p_2\sqrt{\frac{eE}2}e^{-\frac{i\pi}4}[(1+\nu)D_{\nu^*-1}(-\tau^*)+D_{\nu^*+1}
(-\tau^*)]\\
0\\
m_{\perp}^2D_{\nu^*}(-\tau^*)\\
p_2\sqrt{\frac{eE}2}e^{-\frac{i\pi}4}[(1+\nu)D_{\nu^*-1}(-\tau^*)
-D_{\nu^*+1}(-\tau^*)]
\end{bmatrix}e^{i\vec p\cdot\vec x},\eqno(88)
$$
$$
{}^+\psi(2,x)=N(2)\begin{bmatrix}
i(1+\nu)[-D_{\nu^*-1}(-\tau^*)+\frac1{\nu}D_{\nu^*+1}(-\tau^*)]\\
0\\
0\\
i(1+\nu)[-D_{\nu^*-1}(-\tau^*)-\frac1{\nu}D_{\nu^*+1}(-\tau^*)]
\end{bmatrix}e^{i\vec p\cdot\vec x},\eqno(89)
$$
$$
{}^+\psi(3,x)=N(3)\begin{bmatrix}
p_1\sqrt{\frac{eE}2}e^{-\frac{i\pi}4}[(1+\nu)D_{\nu^*-1}(-\tau^*) +
D_{\nu^*+1}(-\tau^*)]\\
(m^2+p_1^2)D_{\nu^*}(-\tau^*)\\
p_1p_2D_{\nu^*}(-\tau^*)\\
p_1\sqrt{\frac{eE}2}e^{-\frac{i\pi}4}[(1+\nu)D_{\nu^*-1}(-\tau^*)-
D_{\nu^*+1}(-\tau^*)]
\end{bmatrix}e^{i\vec p\cdot\vec x}\eqno(90)
$$
In eqs.(85-90) the states $\psi(i,x)$ are characterized by $p_1,p_2,p_3,i$;
$\nu$ and $\lambda$ are given in (7).

We note here that the transition current $J^0({}^+\psi',{}^+\psi)$ in terms
of ${}^+c$ has the same form as          $J^0({}_+\psi',{}_+\psi)$ in  terms
of ${}_+c$, see (82). Similar statement is true for negative-frequency states.
 Taking into account that $\nu+1=-\nu^*$, see (7), we have from (19)
$$
{}^+c'_-{}^*\;{}^+c_+={}_+c'_-{}^*\;{}_+c_+={}^-c'_-{}^*\;{}^-c_+=
{}_-c'_-{}^*\;{}_-c_+. \eqno(90a)
$$
  So
$$
J^0({}_+\psi(i,x),{}_+\psi(j,x))=
       J^0({}^+\psi(i,x),{}^+\psi(j,x))\propto\delta_{i,j}, \eqno(91)
$$
and
$$
J^0({}_-\psi(i,x),{}_-\psi(j,x))=
 J^0({}^-\psi(i,x),{}^-\psi(j,x))=-J^0({}_+\psi(i,x),{}_+\psi(j,x)). \eqno(91a)
$$

As earlier we shall focus our attention on differences from scalar case in
expression similar to (53). The proper-time representation of propagator for
scalar particle is [12]:
\setcounter{equation}{91}
\begin{gather}
G(x,x')_{spin\,0}=\frac{eE}{(4\pi)^2}\exp[\frac i2eE(t+t')z_3]\times\\  \notag
\int\limits_0^{\infty}\frac{ds}{s\sinh eEs}\exp[-ism^2+\frac i{4s}(z_1^2+z_2^2)
+\frac{i}{4}eE(z_3^2-z_0^2)\coth eEs].
\end{gather}
It can be derived similarly to the magnetic case, but the role of eq. (52)
plays the relation [12, 6]
$$
\sqrt2\int\limits_0^{\infty}\frac{d\theta}{\sqrt{\sinh2\theta}}
\exp\left\{-i2\varkappa\theta-\frac i8[\frac{(T+T')^2}{\coth\theta}+
\frac{(T-T')^2}{\tanh\theta}]\right\}=
$$
$$
=\Gamma\left(i\varkappa+\frac12\right)\left\{\begin{array}{cc}
D_{-i\varkappa-\frac12}(\chi)D_{-i\varkappa-\frac12}(-\chi'),\quad T>T'\\
D_{-i\varkappa-\frac12}(-\chi)D_{-i\varkappa-\frac12}(\chi'),\quad T<T'.
\end{array}\right.  \eqno(93)
$$
Here
\setcounter{equation}{93}
\begin{gather}
\theta=eEs,\quad T=\sqrt{2eE}(t-\frac{p_3}{eE}),\quad T'=\sqrt{2eE}(t'-
\frac{p_3}{eE}),\\ \notag
\chi=-\tau^*=e^{\frac{i\pi}4}T,\quad\chi'=e^{\frac{i\pi}4}T',\quad\varkappa=
\frac{\lambda}2=\frac{m_{\perp}^2}{2eE}.
\end{gather}
The lower line on the r.h.s. of (93) can be obtained from the upper line
by substitution $T\leftrightarrow T'$. As seen from the l.h.s. of (93)
this does not change the value of (93), cf. with remark after eq. (53).

By analogy with magnetic case we expect the appearance of factors
$e^{\pm2\theta}$ in the integrand of (93), cf. with eqs. (73), (60-61).
To make the insertion possible, we have to rotate clockwise the integration
 contour by some angle. This is in line with Vanyashin-Terentyev approach [3].
In this way by substitution $\varkappa\to\varkappa+i$ we get from (93)
$$
\sqrt2\int\limits_C\frac{d\theta}{\sqrt{\sinh2\theta}}
\exp\left\{-i2\varkappa\theta+2\theta-\frac i8[\frac{(T+T')^2}{\coth\theta}+
\frac{(T-T')^2}{\tanh\theta}]\right\}=
$$
$$
=\Gamma\left(i\varkappa-\frac12\right)\left\{\begin{array}{cc}
D_{-i\varkappa+\frac12}(\chi)D_{-i\varkappa+\frac12}(-\chi'),\quad T>T'\\
D_{-i\varkappa+\frac12}(-\chi)D_{-i\varkappa+\frac12}(\chi'),\quad T<T'.
\end{array}\right.  \eqno(95)
$$
Similarly, substituting $\varkappa\to\varkappa-i$ in (93) we get
$$
\sqrt2\int\limits_0^{\infty}\frac{d\theta}{\sqrt{\sinh2\theta}}
\exp\left\{-i2\varkappa\theta-2\theta-\frac i8[\frac{(T+T')^2}{\coth\theta}+
\frac{(T-T')^2}{\tanh\theta}]\right\}=
$$
$$
=\Gamma\left(i\varkappa+\frac32\right)\left\{\begin{array}{cc}
D_{-i\varkappa-\frac32}(\chi)D_{-i\varkappa-\frac32}(-\chi'),\quad T>T'\\
D_{-i\varkappa-\frac32}(-\chi)D_{-i\varkappa-\frac32}(\chi'),\quad T<T'.
\end{array}\right.  \eqno(96)
$$

Integration over $p_3$, contained in the sum over $n$ in (80-81) gives
(T, T' are function of $p_3$, see (94))
\setcounter{equation}{96}
\begin{gather}
\int\limits_{-\infty}^{\infty}\frac{dp_3}{2\pi}\exp[ip_3z_3-\frac i8
(T+T')^2\tanh\theta]=\\  \notag
 \frac12e^{-\frac{i\pi}4}\sqrt{\frac{eE\coth\theta}{\pi}}
\exp\left\{\frac{iz_3^2eE}{4\tanh\theta}+\frac{ieEz_3(t+t')}2\right\},
\quad z_3=x_3-x'_3.
\end{gather}
Further calculations leading to (92) are similar to magnetic case.

Now we look for differences from scalar case. First, we rewrite relations
(67-68) between parabolic cylinder functions for present case:
\setcounter{equation}{97}
\begin{gather}
\left(\frac d{d{\tau'}^*}+\frac{{\tau'}^*}2\right)D_{\nu^*}({\tau'}^*)=
\nu^*D_{\nu^*-1}({\tau'}^*),\\  \notag
\left(\frac d{d{\tau'}^*}-\frac{{\tau'}^*}2\right)D_{\nu^*}({\tau'}^*)=
-D_{\nu^*+1}({\tau'}^*), \notag
\end{gather}
\setcounter{equation}{98}
\begin{gather}
2\frac d{d{\tau'}^*}D_{\nu^*}({\tau'}^*)=\nu^*D_{\nu^*-1}({\tau'}^*)-
D_{\nu^*+1}({\tau'}^*),\\  \notag
{\tau'}^*D_{\nu^*}({\tau'}^*)=\nu^*D_{\nu^*-1}({\tau'}^*)+
D_{\nu^*+1}({\tau'}^*).
\end{gather}
Other necessary relations are obtained from these by substitution
${\tau'}^*\to-\tau^*$.

Now taking into account that
$$
c_{1n}=\frac{\sqrt{2\pi}}{\Gamma(-i\varkappa+\frac12)}\exp[-\frac{\pi\varkappa}2+
\frac{i\pi}4],\quad \frac i{c_{1n}^*}N_0^2=\frac{\exp[\frac{3i\pi}4]}
{2\sqrt{\pi eE}}\Gamma(i\varkappa+\frac12),\eqno(100)
$$
we can write the propagator in the form
$$
G^{\mu\nu}(x,x')=\frac{\exp[\frac{3i\pi}4]}{2\sqrt{\pi eE}}\int\frac{d^3p}
{(2\pi)^3}a^{\mu\nu}(x,x')e^{i(\vec p,\vec x-\vec x')}.\eqno(101)
$$
The scalar particle propagator can be obtained from the r.h.s. of (101), if
we substitute $a^{\mu\nu}(x,x')$ by the expression (93).
As an example we calculate now $a^{33}(x,x')$. For $t>t'$ we have
$$
a^{33}(x,x')\propto\sum_{i=1}^3{}^+\psi^3(i,x)\,{}_+\psi^{3*}(i,x').
\eqno(102)
$$
The first term in the sum is
$$
{}^+\psi^3(1,x)\,{}_+\psi^{3*}(1,x')\propto-\frac{ieE}2\tau^*D_{\nu^*}(-\tau^*)
{\tau'}^*D_{\nu^*}({\tau'}^*)\frac{p_2^2}{m_{\perp}^2(m^2+p_1^2)}.\eqno(103)
$$
We have used here the second equation in (99) and the one obtained from it by
substitution ${\tau'}^*\to-\tau^*$. Similarly,
$$
{}^+\psi^3(3,x)\,{}_+\psi^{3*}(3,x')\propto-\frac{ieE}2\tau^*
D_{\nu^*}(-\tau^*){\tau'}^*D_{\nu^*}({\tau'}^*)\frac{p_1^2}{m^2(m^2+p_1^2)}
.\eqno(104)
$$
Summing (103) and (104), we get
$$
-\frac{ieE}2\tau^*D_{\nu^*}(-\tau*){\tau'}^*D_{\nu^*}({\tau'}^*)
[\frac{p_2^2}{m_{\perp}^2(m^2+p_1^2)}+\frac{p_1^2}{m^2(m^2+p_1^2)}]\eqno(105)
$$
The expression in square brackets can be simplified:
$$
\frac1{m^2+p_1^2}(\frac{p_2^2}{m_{\perp}^2}+\frac{p_1^2}{m^2})=\frac1{m^2}-
\frac1{m_{\perp}^2}.\eqno(106)
$$
The undesirable factor $(m^2+p_1^2)^{-1}$, present in (103) and (104),
disappears in the sum (105). The first term on the r.h.s. of (106) gives the
following contribution to (105):
$$
-\frac{ieE}{2m^2}\tau^*D_{\nu^*}(-\tau^*){\tau'}^*D_{\nu^*}({\tau'}^*)=
\frac1{m^2}(p_3-eEt)(p_3-eEt')D_{\nu^*}(-\tau^*)D_{\nu^*}({\tau'}^*).\eqno(107)
$$
This is already the desired form.
Now we rewrite the contribution
 from second term on the r.h.s.
of (106) to (105) in the initial form  (i.e. before using second
 equation in (99)):
\setcounter{equation}{107}
\begin{gather}
\frac{ieE}{2m_{\perp}^2}[-(1+\nu)^2D_{\nu^*-1}(-\tau^*)D_{\nu^*-1}({\tau'}^*)
+(1+\nu)D_{\nu^*+1}(-\tau^*)D_{\nu^*-1}({\tau'}^*)+\\    \notag
+(1+\nu)D_{\nu^*-1}(-\tau^*)D_{\nu^*+1}({\tau'}^*)-
D_{\nu+1}(-\tau^*)D_{\nu^*+1}({\tau'}^*)].
\end{gather}
This expression still contains undesirable factor $\frac1{m_{\perp}^2}$.
But we must take into account the contribution from term with $i=2$ in (102):
\setcounter{equation}{108}
\begin{gather}
{}^+\psi^3(2,x)\,{}_+\psi^{3*}(2,x')\propto\frac{ieE}{2m_{\perp}^2}(1+\nu)
[-D_{\nu^*-1}(-\tau^*)D_{\nu^*+1}({\tau'}^*)-\\      \notag
\frac1{\nu}D_{\nu^*+1}(-\tau*)D_{\nu^*+1}({\tau'}^*)-\nu D_{\nu^*-1}(-\tau^*)
D_{\nu^*-1}({\tau'}^*)-D_{\nu^*+1}(-\tau^*)D_{\nu^*-1}({\tau'}^*)].
\end{gather}
It is easy to see that in the sum of (108) and (109) undesirable terms
are cancelled and unpleasant denominator $m_{\perp}^2=-ieE(1+2\nu)$ disappears:
$$
(108)+(109)=\frac12[(1+\nu)D_{\nu^*-1}(-\tau^*)D_{\nu^*-1}({\tau'}^*)+
\frac1{\nu}D_{\nu^*+1}(-\tau^*)D_{\nu^*+1}({\tau'}^*)].\eqno(110)
$$
Thus $a^{33}(x,x')$ is given by the sum of expressions (107) and (110).
The first term on the r.h.s. of (110) is used in (96) and the second term in
(95). In the same manner we find all other $a^{\mu\nu}(x,x')$. Similarly
to the magnetic case we have
\setcounter{equation}{110}
\begin{gather}
G^{\mu\nu}(x,x')=\frac{eE}{(4\pi)^2}\int\limits_C
\frac{ds}{s\sinh eEs}A^{\mu\nu}\exp[\frac{ieE}2z_3(t+t')]\times\\  \notag
\exp[-ism^2+\frac i{4s}(z_1^2+z_2^2)+\frac i{4s}(z_3^2-z_0^2)eE\coth eEs].
\end{gather}
Here $A^{\mu\nu}$ has the form (71), but the vector-potential is
$A_{\mu}(x)=-\delta_{\mu3}Et$. The nonzero $B^{\mu\nu}$ are
$$
B^{11}=B^{22}=1,\quad B^{33}=-B^{00}=\cosh 2eEs,\quad
 B^{30}=-B^{03}=\sinh 2eEs.  \eqno(112)
$$
 We see that electric field dresses $B^{\mu\nu}$ with $\mu,\nu=3,0$.

Moving on to the case $t<t'$, we note that according to (19)
$$
{}_-c_{\pm}=-{}^+c_{\pm},\quad {}^-c_{\pm}=-{}_+c_{\pm}.\eqno(112a)
$$
It follows from here that ${}_-\psi \;({}^-\psi)$ is obtained from
${}^+\psi \;({}_+\psi)$ by changing sign of arguments of parabolic cylinder
functions and sign of $\psi^0$ and $\psi^3$. The overall change of sign of
$\psi(2,x)$ does not tell on corresponding term in (102). In $\psi(1,x)$
 and $\psi(3,x)$ changing sign of
 $\psi^0$, $\psi^3$ and arguments $\tau^*$, $\tau'{}^*$ is equivalent
to changing sign of only arguments  $\tau^*$, $\tau'{}^*$, when
 $\psi^0$ and $\psi^3$ are expressed through the left hand sides of (99).
Now, as expected, it follows from (93-96) that $G^{\mu\nu}(x,x')$ retains the
same form (111) for $t<t'$.

\section{Propagator of vector boson in constant electromagnetic field}

After we have considered separately the magnetic and electric fields, the
building up of the propagator of vector boson in both fields meets with no
new problems. We take vector-potential in the form
$$
A_{\mu}(x)=\delta_{\mu2}Hx_1-\delta_{\mu3}Et. \eqno(113)
$$
The transition current between states ${}_+\psi'$ and ${}_+\psi$ is
\setcounter{equation}{113}
\begin{gather}
J^0({}_+\psi',{}_+\psi)=2\{[{c'_1}^*c_1D_{n-1}^2(\zeta)+{c'_2}^*c_2
D_{n+1}^2(\zeta)]D_{\nu^*}(\tau^*)i\buildrel\leftrightarrow\over{\frac d{dt}}
D_{\nu}(\tau)- \\ \notag
[{c'_-}^*\;c_+%
D_{\nu^*-1}(\tau^*)%
 i\buildrel\leftrightarrow\over{\frac d{dt}}D_{\nu+1}(\tau)
+{c'_+}^*\;c_-
D_{\nu^*+1}(\tau^*)i\buildrel\leftrightarrow\over{\frac d{dt}}D_{\nu-1}(\tau)]
D_n^2(\zeta)\}.
\end{gather}
Taking into account (84) and integrating over $x_1$ we get
$$
\int\limits_{-\infty}^{\infty}dx_1J^0({}_+\psi',{}_+\psi)
=n!\sqrt{\frac{2\pi E}H}2
e^{\frac{\pi\varkappa}2}[\frac1n{c'_1}^*c_1+(1+n){c'_2}^*c_2 +
i({c'_-}^*c_+-{c'_+}^*c_-)] \eqno(115)
$$
The constraint has the form
$$
\sqrt{2eH}[(1+n)c_2-c_1]+\sqrt{2eE})e^{-\frac{i\pi}4}[{}_+c_-
-(1+\nu)\,{}_+c_+]=0.
\eqno(116)
$$
Using (115) and (116) we find the ${}_+\psi$ polarization states (in the
following the factor $e^{i(p_2x_2+p_3x_3)}$  is dropped for brevity):
$$
{}_+\psi(1,x)=N(1)\begin{bmatrix}
(1+n)\sqrt{e^2EH}e^{\frac{i\pi}4}[D_{\nu+1}(\tau)-\nu D_{\nu-1}(\tau)]D_n(\zeta)
\\
im_{\perp}^2D_{\nu}(\tau)D_{n+1}(\zeta)\\
 m_{\perp}^2D_{\nu}(\tau)D_{n+1}(\zeta)\\
(1+n)\sqrt{e^2EH}e^{\frac{i\pi}4}[D_{\nu+1}(\tau)+\nu D_{\nu-1}(\tau)]D_n(\zeta)
\end{bmatrix},\;  \eqno(117)
$$
$$
N(i)=n_iN_0,\;N_0=\left(\frac H{2\pi E}\right)^{\frac14}
\frac{\exp[-\frac{\pi\varkappa}
4]}{\sqrt{n!}},\;
$$
$$
 n_1=\frac1{\sqrt{2m_{\perp}^2(m^2+eHn)(1+n)}},\eqno(118)
$$
$$
  {}_+\psi(2,x)=N(2)\begin{bmatrix}
[D_{\nu+1}(\tau)+(1+\nu)D_{\nu-1}(\tau)]D_n(\zeta)\\
0\\
0\\

[D_{\nu+1}(\tau)-(1+\nu)D_{\nu-1}(\tau)]D_n(\zeta)
\end{bmatrix},
\;n_2=\sqrt{\frac{eE}{2m_{\perp}^2}},\eqno(119)
$$
$$
{}_+\psi(3,x)=N(3)\begin{bmatrix}
\sqrt{e^2EH}[D_{\nu+1}(\tau)-\nu D_{\nu-1}(\tau)]D_n(\zeta)\\
e^{\frac{i\pi}4}D_{\nu}(\tau)[-(m^2+eHn)D_{n-1}(\zeta)+eHD_{n+1}(\zeta)]\\
e^{-\frac{i\pi}4}D_{\nu}(\tau)[(m^2+eHn)D_{n-1}(\zeta)+eHD_{n+1}(\zeta)]\\
\sqrt{e^2EH}[D_{\nu+1}(\tau)+\nu D_{\nu-1}(\tau)]D_n(\zeta)\\
\end{bmatrix}, \eqno(120)
$$
$$
n_3=\sqrt{\frac n{2m^2(m^2+eHn)}}, \eqno(121)
$$

To obtain polarization states of ${}^+\psi$ (or ${}_-\psi, {}^-\psi$ for
that matter) we use again (19) (cf. with (88-90), (47-50)). Then we get
$$
{}^+\psi(1,x)=
$$
$$
N(1)\begin{bmatrix}
(1+n)\sqrt{e^2EH}e^{-\frac{i\pi}4}[(1+\nu)D_{\nu^*-1}(-\tau^*)+
D_{\nu^*+1}(-\tau^*)]D_n(\zeta)\\
im_{\perp}^2D_{\nu^*}(-\tau^*)D_{n+1}(\zeta) \\
m_{\perp}^2D_{\nu^*}(-\tau^*)D_{n+1}(\zeta) \\
(1+n)\sqrt{e^2EH}e^{-\frac{i\pi}4}[(1+\nu)D_{\nu^*-1}(-\tau^*)-
D_{\nu^*+1}(-\tau^*)]D_n(\zeta)\\
\end{bmatrix}, \eqno(122)
$$
$$
{}^+\psi(2,x)=N(2)\begin{bmatrix}
i(1+\nu)[-D_{\nu^*-1}(-\tau^*)+\frac1{\nu}D_{\nu^*+1}(-\tau^*)]D_n(\zeta)\\
0\\
0\\
i(1+\nu)[-D_{\nu^*-1}(-\tau^*)-\frac1{\nu}D_{\nu^*+1}(-\tau^*)]D_n(\zeta)\\
\end{bmatrix}, \eqno(123)
$$
$$
{}^+\psi(3,x)=N(3)\begin{bmatrix}
-i\sqrt{e^2EH}[-\nu^*D_{\nu^*-1}(-\tau^*)+D_{\nu^*+1}(-\tau^*)]D_n(\zeta)\\
e^{\frac{i\pi}4}D_{\nu^*}(-\tau^*)[-(m^2+eHn)D_{n-1}(\zeta)+eHD_{n+1}(\zeta)]\\
e^{-\frac{i\pi}4}D_{\nu^*}(-\tau^*)[(m^2+eHn)D_{n-1}(\zeta)+eHD_{n+1}(\zeta)]\\
 -i\sqrt{e^2EH}[-\nu^*D_{\nu^*-1}(-\tau^*)-D_{\nu^*+1}(-\tau^*)]D_n(\zeta)\\
\end{bmatrix}. \eqno(124)
$$
The first and the fourth lines on the r.h.s. of (122) and (124) can be written
in more compact form with the aid of relations obtainable from (99) by
substitution ${\tau'}^*\to-\tau^*$.

Further calculations are quite similar to those in Sections (5) and (6). The
result is evident, of course, beforehand: now $A^{\mu\nu}$ is given by (71)
with vector-potential (113) and all nonzero $B^{\mu\nu}$ are "dressed", see
(73) and (112). The propagator of scalar particle has the form
\setcounter{equation}{124}
\begin{gather}
G_{spin\,0}(x,x')=\frac{e^2EH}{(4\pi)^2}\int\limits_0^{\infty}
\frac{ds}{\sinh eEs\;\sin eHs}\exp\{-ism^2+\\ \notag
\frac i4[(z_1^2+z_2^2)eH\cot eHs+(z_3^2-z_0^2)eE\coth eEs]+ \\ \notag
\frac i2[eEz_3(t+t')-eHz_2(x_1+x'_1)]\},\\ \notag
z_{\mu}=x_{\mu}-x'_{\mu}.
\end{gather}
This expression is in agreement with Ritus calculations [10-11]. The presence
of phase factor $e^{-\frac{i\pi}2}$ in his formulas is due to difference
in definition of propagator. We note also that (125) is symmetric in $t,t'$ and
$G_{spin\,0}(x,x',e)=
G_{spin\,0}(x',x,-e)$. Thus
\setcounter{equation}{125}
\begin{gather}
G^{\mu\nu}(x,x')=\frac{e^2EH}{(4\pi)^2}\int\limits_0^{\infty}
\frac{ds}{\sinh eEs\;\sin eHs}A^{\mu\nu}\exp\{-ism^2+\\ \notag
\frac i4[(z_1^2+z_2^2)eH\cot eHs+(z_3^2-z_0^2)eE\coth eEs]+ \\ \notag
\frac i2[eEz_3(t+t')-eHz_2(x_1+x'_1)]\}.
\end{gather}

\noindent{\bf Acknowledgements.}
 The author is grateful to  V.S.Vanyashin, V.Incera, and E.Ferrer
 for stimulating discussions. Special thanks are to V.I.Ritus for suggesting
valuable improvements in the manuscript.
 This work was supported in part by RFFR
grant No. 00-15-96566.


\begin{thebibliography}{7}
\bibitem{1}
V.V.Skalozub, Particles and Nuclei, V. 16, p. 1005 (1985).
\bibitem{2}
V.M.Mostepanenko, V.M.Frolov, V.A.Sheluto, Yad. Fiz. V. 37,
 p. 1261 (1983).
\bibitem{3}
 V.S.Vanyashin, M.V.Terentyev. Zh. Eksp. Teor. Fiz. V. 48, p. 565 (1965).
\bibitem{4}
M.S.Marinov, V.S.Popov, Yad. Fiz. V. 15, p. 1271 (1972).
\bibitem{5}
A.A.Grib, S.G.Mamaev, V.M.Mostepanenko, Vacuum quantum effects in strong fields.
Moscow, 1988.
\bibitem{6}
A.I.Nikishov, Trudy FIAN, V. 111, p. 152, Nauka, Moscow (1979).
                                              (Journal of Soviet
 Laser Research, V. 6, Nov-Dec. p. 619 (1985)).
\bibitem{7}
Erd\'elyi A. et al, Higher Transcendental Functions,
V. 2, Mc Grow-Hill Book Co. 1953.
\bibitem{8}
V.B.Berestetskii, E.M.Lifshitz, L.P.Pitaevskii, Quantum electrodynamics,
Moscow, 1980.
\bibitem{9}
J.Schwinger, Phys. Rev. V. 82, p. 661 (1951).
\bibitem{10}
V.I.Ritus, ZhETP, V. 75, p. 1560 (1978). (Sov. Phys. JETP, V. 48, p. 788 (1978));
 Trudy FIAN, V. 111, p. 134, Nauka, Moscow, (1979). (Journal of
Soviet Laser Research, V. 6, No 5, Sept-Oct., p. 497 (1986).)
\bibitem{11}
A.I.Nikishov, V.I.Ritus, Trudy FIAN, V. 168, Nauka, Moscow (1986).
 (Issues in
Intense-Field Quantum Electrodynamics. Ed. by V.L.Ginzburg. Nova Science
Publishers. Inc. Commack (1987).)
\bibitem{12}
A.I.Nikishov, ZhETP, V. 57, p. 1210 (1969). (Sov.Phys. JETP,
 V. 30, p. 660 (1970)).
\end{thebibliography}
\end{document}